\documentclass[epsf]{revtex4}

\usepackage{graphicx}

\begin{document}

\title{Causality in Inflationary Universes with Positive Spatial Curvature}
\author{G. F. R. Ellis$^{1,2}$, P. McEwan$^1$, W. Stoeger$^3$, 
and P. Dunsby$^1$}

\address{$1$ Department of Applied Mathematics, University of Cape Town, 
Rondebosch 7700, Cape Town, South Africa.}
\address{$2$ Erwin Schroedinger Institute, Vienna, Austria.}
\address{$3$ Vatican Observatory, Tucson, Arizona, USA.}

\date{\today}

\begin{abstract}
We show that in the case of positively-curved Friedmann-Lema\^{\i}tre
universes $(k=+1)$, an inflationary period in the early universe will
for most initial conditions not solve the horizon problem, no matter
how long inflation lasts. It will only do so for cases where inflation
starts in an almost static state, corresponding to an extremely high
value of $\Omega_{\Lambda}$, $\Omega_{\Lambda} \gg 1$, at the
beginning of inflation. For smaller values, it is not possible
to solve the horizon problem because the relevant integral asymptotes
to a finite value (as happens also in the de Sitter universe in a
$k=+1$ frame). Thus, for these cases, the causal problems associated
with the near-isotropy of the Cosmic Background Radiation have to be
solved already in the Planck era. Furthermore both compact space
sections and event horizons will exist in these universes even if the
present cosmological constant dies away in the far future, raising
potential problems for M-theory as a theory of gravity.
\end{abstract}

\maketitle
\section{Inflation Causality with Positive Spatial Curvature}
Recent measurements of a second and third peak in the cosmic blackbody
background radiation (CBR) anisotropy spectrum \cite{Ref2} together with
supernova data \cite{Ref6} suggest best-fit inflationary universe models 
\cite{Ref1} with a non-zero cosmological constant and sufficient matter to
make it almost flat ($\Omega _{\Lambda _0}\approx 0.7,\,\Omega _{m_0}\approx
0.3,\,\Omega _0\approx 1)$  \cite{Ref3}. While the set of models compatible
with the data include those with flat spatial sections $(k=0)$ and so with a
critical total effective energy density ($\Omega _0=1$ exactly), they also
include positive spatial curvature $(k=+1:\Omega _0>1\;)$ models and
negative curvature $(k=-1:\Omega _0<1)$ ones, with a weak implication that
the best-fit models have positive curvature\emph{\ }\cite{Ref3}. It should
be noted that while inflation is taken to predict that the universe is very
close to flat at the present time, it does \emph{not} imply that the spatial
sections are \emph{exactly} flat; indeed that case is infinitely improbable,
and neither inflation nor any other known physical process is able to
specify that curvature, nor dynamically change it from its initial value 
\cite{ell_ch}. Thus there is no reason to believe on the basis of
inflationary dynamics that $k=0,$ and it is certainly worth exploring the
properties of positive-curvature inflationary models \cite{pos}, which can
be taken to be marginally indicated by present observations.

We have shown \cite{Paper_1} that in such universes, basically because these
positive curvature solutions (unlike those in the case $k=0$) are not
scale-invariant and have to be compatible with the present-day Cosmic
Background Radiation (CBR) energy density, there are limits to the
numbers of e-foldings that are possible, independent of the pre-inflationary
dynamics. One might suspect that this would imply a limit to the ability of
these models to solve the  \emph{Horizon Problem} \footnote{ 
Note that the `horizon' referred to here is a \emph{particle horizon},
dependent on properties of the very early universe, rather than an \emph{\
event horizon,} dependent on properties of the very late universe. Both
should be distinguished form the \emph{Hubble scale} $H^{-1}c$, often also
called the {\it horizon}, which is a local quantity that is not directly
dependent on either the very early universe or the very late universe.} -
the causality issues raised by the very high degree of isotropy of the CBR 
\cite{Ref1}. We show that this is indeed the case, but - surprisingly - it
is not directly related to the limit to the possible number of e-foldings, but
rather to the magnitude of the dominant vacuum energy (cosmological constant),
and therefore to the effective initial time, at the beginning of inflation.

The issue here is that points of emission of this radiation on the surface
of last scattering (LSS) are causally disconnected in a standard Hot Big
Bang model, i.e. a Friedman-Lema\^{\i}tre (FL) universe model that is
matter dominated at recent times and radiation-dominated at early times,
because, irrespective of the value of $\emph{k}$, they lie beyond each
other's particle horizons. Hence in this case there can be no causal
explanation of why conditions are so similar at this surface as to lead to
an almost isotropic CBR at the present time; indeed the radiation detected 
by the COBE satellite and the BOOMERANG balloon would have originated from
matter in  many different regions causally disconnected from each other at the
time of emission of that radiation \cite{Ref4}, \cite{Ref4a}. In a 
$k=0$ FL universe, a period of exponential expansion (inflation) in the early
universe solves this problem by increasing the particle horizon size at last
scattering many-fold. This leads to the claim \cite{Ref1} that the Horizon
Problem is solved in inflationary universes, thus allowing a causal
explanation of why the universe is as homogeneous as it is . One should note
here that in the standard lore of inflation, the horizon - not the
particle horizon but the Hubble scale $H^{-1}c$ - is considered constant
during inflation, and this plays a crucial role in structure formation
scenarios; however that length scale has only an indirect relation to
causality in terms of propagation of effects at speeds less than or equal to
the speed of light.

The usual assumption is that inflation solves the horizon problem even if $ 
\Omega _0$ is not exactly unity, i.e. even if the spatial sections are not
exactly flat. This claim is not as straightforward as it seems. We consider
here positive curvature models $(k=+1)$, and show that their causal horizons
are quite different from those in $k=0$ and $k=-1$ models, even if they are
extremely close to being flat at the present time. Indeed, however much
inflation takes place and irrespective of how close to flat the model is at
the present time, there are many positively curved models where inflation
does not solve the horizon problem. In fact, there are two separate horizon
issues in $k=+1$ models. The first is whether or not the distance to the
particle horizon becomes equal to or larger than the radius of the spatial
hypersurfaces by decoupling, thus causally connecting the entire universe. The
second issue is the traditional horizon problem: Is the size of the particle
horizon at decoupling larger than or equal to  the size of the visual horizon
now? If it is not (that is, if the particle horizon is smaller than the visual
horizon), then the horizon problem is not solved. There are some extreme cases
in which the particle horizon embraces the entire universe after inflation
(the first issue) -- this automatically solves the less demanding horizon
problem as well. The most realistic of these depend on having an extremely
large $\Omega_{\Lambda} \gg 1$ at the beginning of inflation. Even if causal
connectedness throughout the entire universe is not achieved, the horizon
problem can be solved if the second criterion is satisfied. But it turns out
again that this only happens if we start inflation with a very, very high
$\Omega_{\Lambda}$, though not as extreme as demanded by total
causal connectedness, other parameters being equal. There will, therefore, be
many $k=+1$ inflationary universes in which the horizon problem cannot be
solved by inflation itself, no matter how many e-foldings are applied. This
will be explained in detail later in the paper.

Furthermore, even if the horizon is reached by photons from every part of
the universe by the time of decoupling, as is possible in the extreme cases
referred to above, there is
still totally insufficient causal contact during inflation to allow physical
processes in that epoch to homogenize the universe by that time. 
In particular, chemical homogeneity then depends on adequate causal
contact being established by the time of nucleosynthesis. In most inflationary
universe models with $k=+1$ that is unachievable. Our calculations are
for the case of a constant vacuum energy during the inflationary era; there
should be no difference for inflation driven by a slow-rolling scalar field,
because at early enough times, the spatial curvature term will dominate the
Friedman equation in these cases also; however power-law inflationary models
could give different answers.

A further point that has recently raised interest is the claim that the
existence of event horizons \cite{Ref4}, \cite{Ref4a} in a FL\ universe
creates problems for string theory (or M-theory) as a fundamental theory of
gravity \cite{Ref5}, because there are then problems in setting data for a
Scattering Matrix. Such horizons occur if the late universe is dominated by
a cosmological constant, as is suggested by current observations of
supernovae in distant galaxies \cite{Ref6}. It has been suggested however
that this problem will go away if that constant is actually variable
(quintessence), and decays away in the far future, so the universe does
not undergo eternal exponential expansion \cite{Ref7}. We point out here
that this resolution of the problem is not possible if $k=+1$, for event
horizons will occur in this case whether there is a cosmological constant or
not, and quintessence will not change that situation; and furthermore
additional problems arise because the spatial sections are compact, so an
infinity where one can set data in the spirit of the `holographic universe'
proposals does not exist in this case. Thus astronomical evidence that the
universe has positive spatial curvature may be evidence against the validity
of M-theory.

Although the evidence is that there is currently a non-zero cosmological
constant, as mentioned above, for simplicity we will consider here mainly
the case of an almost-flat $k=+1$ universe with vanishing cosmological
constant after the end of inflation. This approximation will not affect the
statements derived concerning causality up to the time of decoupling, but
will make a small difference to estimates of apparent angles. We use a
simple multi-stage model with exponential inflation, rather than a
continuous model of the change of the effective equation of state and a
dynamic scalar field. A further paper \cite{Paper_2} will improve on these
approximations and give more details of the numerical results.

\section{Basic Equations}
\subsection{Geometry and Light Propagation}

The FL cosmological model considered here is described by a
Robertson-Walker metric for $k=+1$:

\begin{equation}
ds^2=-c^2dt^2+S^2(t)d\sigma ^2,\;\;d\sigma ^2=dr^2+\sin ^2r\;(d\theta
^2+\sin ^2\theta d\phi ^2)  \label{Metric}
\end{equation}
in comoving coordinates $(t,r,\theta ,\phi ),$ so the 4-velocity of
fundamental observers is $u^a=\delta _0^a.$ Here $c$ is the speed of light, $ 
r$ is dimensionless, $t$ has dimensions of time, and $S(t)$ has dimensions
of distance. The scale factor S(t) is normalized so that the spatial metric
has unit spatial curvature at the time $t_{*}$ when $S(t_{*})=1$ (see e.g. 
\cite{Wein},\cite{brazil}). The Hubble Parameter is $H(t)=\dot{S}(t)/S(t),$
with dimensions of (time)$^{-1}$ and present value $H_0=100h\,\,km/sec/Mpc$;
the dimensionless quantity $h$ probably lies in the range $0.5<h<0.7$. The
spatial sections are closed at $r-$coordinate value increment $2\pi ;$ that
is, $P=(t,r-\pi ,\theta ,\phi )$ and $P^{\prime }=(t,r+\pi ,\theta ,\phi )$
are necessarily the same point, for arbitrary values of $t,r,\theta ,\phi ,$
and wherever the origin of coordinates is chosen. Thus the spatial distance
of any point from any other point cannot exceed the equivalent of an
$r$-increment of $\pi ,$ which at time $t$ is equal to a distance $\pi S(t)$.

We need to determine light propagation on radial null geodesics ($d\theta
=d\phi =0=ds^2)$ connecting different fundamental world lines. Light emitted
by a comoving observer $A$ $(r=r_A)$\ at time $t_A$ and received by a
comoving observer $B$ $(r=r_B)$ at time $t_B$ obeys

\begin{equation}
\Psi (A,B)\equiv c\int_{S_A}^{S_B}\frac{dS}{S\dot{S}}=\int_{t_A}^{t_B}\frac{
cdt}{S(t)}=r_B-r_A.  \label{Psi}
\end{equation}
This integral gives the comoving distance between $A$ and $B,$ normalized to
the actual distance at the time $t_{*}$ when $S(t_{*})=1,$ in terms of the
conformal time $\tau $ used in the usual conformal diagrams of light
propagation in FL\ universes \cite{Ref4a}. Note that in a chain of such
observations, $\Psi (A,C)=\Psi (A,B)+\Psi (B,C).$ The physical distance
between $A$ and $B$ at some reference time $t_R$ is

\begin{equation}
D(A,B)=S(t_R)\Psi (A,B).  \label{Hd}
\end{equation}

\subsection{Dynamic Equations}

The integral in (\ref{Psi}) is determined dynamically by the value of $\dot{
S }$ determined by the Friedman equation for $k=+1$:

\begin{equation}
\left( \frac{H(t)}c\right) ^2=\frac{\kappa \mu (t)}3+\frac \Lambda 3-\frac
1{S(t)^2}  \label{Fried}
\end{equation}
where $\kappa $ is the gravitational constant in appropriate units and $ 
\Lambda $ the cosmological constant (see e.g. \cite{Wein},\cite{brazil}).
The way this works out in practice is determined by the matter content of
the universe, whose total energy density $\mu (t)$ and pressure $p(t)$
necessarily obey the conservation equation

\begin{equation}
\dot{\mu}(t)+\left( \mu (t)+p(t)/c^2\right) 3H(t)=0.  \label{Cons}
\end{equation}
The nature of the matter is determined by the equation of state relating $ 
p(t)$ and $\mu (t);$ we will describe this in terms of a parameter $\gamma
(t)$ defined by
\begin{equation}
p(t)=c^2\left( \gamma (t)-1\right) \mu (t),\;\gamma (t)\in [0,2].\ 
\label{EqnState}
\end{equation}
During major epochs of the universe's history, the matter behaviour is
well-described by this relation with $\gamma $ a constant (but with that
constant different at various distant dynamical epochs). In particular, $ 
\gamma =1$ represents pressure free matter (baryonic matter), $\gamma
=\frac 43$ represents radiation (or relativistic matter), and $\gamma =0\ $
gives an effective cosmological constant of magnitude $\Lambda =\kappa \mu $
(by equation (\ref{Cons}), $\mu $ will then be unchanging in time). In
general, $\mu (t)$ will be a sum of such components. During a \emph{\
cosmological constant-dominated era}, i.e. when $\Lambda >0$ and we can
ignore matter and radiation in (\ref{Fried}), with a suitable choice of the
origin of time we obtain the simple collapsing and re-expanding solution: 
\begin{equation}
S(t)=\frac c\lambda \cosh \lambda t,\;\lambda \equiv c\sqrt{\frac \Lambda 3\ 
}.  \label{CC}
\end{equation}
This is of course just the de Sitter universe represented as a
Robertson-Walker space-time with positively-curved space sections \cite{Schr}
, and can be used to represent an inflationary era for universe models with $ 
k=+1\;$ if we restrict ourselves to the expanding epoch: 
\begin{equation}
\;t\geq t_i\geq 0\Rightarrow \exp (\lambda t)\;=\frac \lambda cS(t)+\sqrt{ 
\frac{\lambda ^2S(t)^2}{c^2}-1}  \label{t_i}
\end{equation}
for some suitable initial time $t_i$.

The density parameter $\Omega (t)$ and associated quantity $\ \Sigma (t)$
are defined by 
\begin{equation}
\Omega (t)\equiv \frac{\kappa \mu (t)}3\left( \frac c{H(t)}\right) ^2>1,\;\
\Sigma (t)\equiv \frac{\Omega (t)-1}{\Omega (t)}\in [0,1).  \label{Omega}
\end{equation}
We can include a cosmological constant in terms of an equivalent energy
density $\kappa \mu _\Lambda =\Lambda ;$ from now on we omit explicit
reference to $\Lambda ,$ assuming it will be represented in this way when
necessary. For each epoch where $\gamma $ is constant, provided $\gamma \neq
2/3$ \footnote{ 
We omit the unphysical case $\gamma =2/3.$}, using (\ref{Fried}) in (\ref
{Psi}) gives 
\begin{equation}
\Psi (\gamma ,A,B)=\int_{\sigma _A}^{\sigma _B}\frac{d\sigma }{\sigma \sqrt{
\sigma ^{2-3\gamma }-1}}  \label{Int}
\end{equation}
where the dimensionless quantities $\sigma $ are defined by 
\begin{equation}
\sigma _A=\frac{S_A}{S_Q}\left( \ \Sigma _Q\right) ^{\frac 1{2-3\gamma
}},\;\sigma _B=\frac{S_B}{S_Q}\left( \Sigma _Q\right) ^{\frac 1{2-3\gamma }},
\label{IntConst}
\end{equation}
and $\Sigma _Q\equiv \Sigma (t_Q)$ is evaluated at some reference point $Q$
in the period of constant $\gamma$ (or possibly one of the end-points).

\subsection{Horizons and Causality}

The distance light travels to reach us receives contributions from different
eras, possibly including the Planck era. Consider zero-rest-mass radiation
traveling towards us on a null geodesic from the origin of the universe, or
at least from the Planck time. Let event\ $A$ be at the end of the Planck
era, with $t=t_{Planck}$ allowing for a radiation-dominated era before 
the start of inflation. Let event $B$ be the start of inflation
(possibly the same as $A$), with $t=t_i$; let event $C$ \ be at end of
inflation, i.e. the start of the radiation dominated era, with $t=t_f$; let
event $D$ be at the end of the radiation dominated era, i.e. the start of
matter the dominated era, with $t=t_{eq}$; we will take this to be the time
of decoupling; and let event $E$ be today, with $t=t_0$. Note that all these
events are located in the expanding domain of the universe (this is
important later in terms of limits on integrals). Thus the comoving \emph{\
particle horizon size }today, representing the causal contact that can have
been attained since the beginning of the universe until today \cite{Ref4}, 
\cite{Ref4a}, is 
\begin{equation}
\Psi _0=\Psi _{Planck}+\Psi _1+\Psi _2+\Psi _3+\Psi _4,  \label{Horiz_t0}
\end{equation}
where $\Psi _1\equiv \Psi (\frac 43,A,B),\;\Psi _2\equiv \Psi (0,B,C),$ $ 
\Psi _3\equiv \Psi (\frac 43,C,D),$ and $\Psi _4\equiv \Psi (1,D,E),$ the
various terms in (\ref{Horiz_t0}) representing the range of causal
connection at the start of inflation (resulting from processes in the Planck
era), and contributions from the initial radiation dominated era, the
inflationary era, the later radiation dominated era, and the matter
dominated era, respectively \footnote{ 
We should in principle add also a late cosmological constant dominated era
by interpolating a point $F$ between $D$ and $E$, so that $\Psi _{4D}\equiv
\Psi (1,D,F)$ is matter dominated and $\Psi _{4E}\equiv \Psi (0,F,E)$ is
cosmological-constant dominated, where $F$ corresponds to a redshift of $ 
z=0.326,$ and $\Psi _4=\Psi _{4D}+\Psi _{4E}$. However we will omit this
extra complication for simplicity; this will not substantially change the
results.}. The corresponding physical distance at the present time $t_0$ is $ 
D_H(t_0)=S(t_0)\Psi _0.$ Furthermore the comoving particle horizon size at
decoupling ($t=t_{eq}$) is 
\begin{equation}
\Psi _H(t_{eq})=\Psi _{Planck}+\Psi _1+\Psi _2+\Psi _3  \label{Horiz_teq}
\end{equation}
and the comoving particle horizon size at the end of inflation ($t=t_f$) is 
\begin{equation}
\Psi _H(t_f)=\Psi _{Planck}+\Psi _1+\Psi _2.  \label{Horiz_tf}
\end{equation}
We have causal connectivity of all particles in the universe at those times
if $\Psi
_H(t_{eq})\geq \pi ,$ $\Psi _H(t_f)\geq \pi $ respectively (note that light
goes in both directions, and we have calculated this distance only for one
direction; that is why the number here is $\pi $ rather than $2\pi ,$ which
is the spatial distance corresponding to spatial closure). The corresponding
physical distances at the LSS, i.e. the corresponding comoving distances as
reflected in the COBE and BOOMERANG\ maps, are

\begin{equation}
D_H(t_{eq})=S(t_{eq})\Psi _H(t_{eq}),\;D_H(t_f)=S(t_{eq})\Psi _H(t_f).
\label{Horiz_Dist}
\end{equation}
These quantities represent the comoving horizon sizes at decoupling and at
the end of inflation respectively, translated into physical distances on the
surface of last scattering. This connectivity depends on that which already
exists as the universe emerges from the Planck era, represented by $\Psi
_{Planck},$ and that gained after the Planck era, represented by the rest of
these expressions. We will for the moment set $\Psi _{Planck}=0\ $ in order
to investigate the causal connectivity attained after the Planck era; we
will return to considering the effect of non-zero $\Psi _{Planck}$ in a
later section.

Finally we note that $\Psi _4$ represents the size of the \emph{visual
horizon }\cite{Horiz}: 
\begin{equation}
\Psi _{VH}(t_{eq})=\Psi _4,  \label{VisHor}
\end{equation}
characterizing the set or particles we can actually have seen by
electromagnetic radiation at any wavelength (it represents the maximum
comoving distance light can have traveled towards us from any object, this
distance being limited by the opaqueness of the universe prior to
decoupling).

\subsection{Joining different eras}

Junction conditions required in joining two eras with different equations of
state are that we must have $S(t)$ and $\dot{S}(t)$ continuous there, thus $ 
H(t)$\ is continuous also. By the Friedman equations this implies in turn
that $\mu (t)$ is continuous, so by its definition $\Omega (t)$ is also
continuous (note that it is $p(t)$ that is discontinuous on spacelike
surfaces of discontinuity). We need to demand, then, that any two of these
quantities are continuous where the equation of state is discontinuous; for
our purposes it will be convenient to take them as $S(t)$ and $\Omega (t).$
Thus in calculating the contributions to $\Psi $, we assume epochs with
constant $\gamma $ joined according to these junction conditions (see \cite
{Paper_2} for details). Note that we can use different time parameters in
each epoch, if that is convenient; all that we requires is that these
junction conditions are satisfied.

\section{Causal Limits in Positive Curvature Models}

One might naively expect that during an inflationary era with at least 60
e-foldings, complete mixing could take place in a universe with closed
spatial sections - causal influences could travel round the universe many
times. However this is not so when $k=+1$, although it is true in spatially
compact universes with $k=0$ and $k=-1$. When $k=+1$, the dynamics of the
universe is importantly different at early times, and consequently the
integral (\ref{Int}) converges, even if there is inflation, to less than the
amount needed to see round the universe many times.

To derive limits on contributions to $\Psi $ in the inflationary, radiation,
and matter eras, we use the following evaluations of (\ref{Int}) for
constant values of $\gamma $. For the matter-dominated era, we set $\gamma
=1 $ and obtain 
\begin{equation}
\Psi _4\equiv \Psi (1,D,E)=\arcsin \left[ \frac{2S_E}{S_R}\Sigma _R-1\right]
-\arcsin \left[ \frac{2S_D}{S_R}\Sigma _R-1\right]\;,  \label{IntDust}
\end{equation}
where $R$ is a reference point in this period, and $S_D<S_E\leq S_R\Sigma _R$
. For the later radiation-dominated era, we set $\gamma =4/3$ and obtain 
\begin{equation}
\Psi _3\equiv \Psi (\frac 43,C,D)=\arcsin \left[ \frac{S_D}{S_Q}\sqrt{\Sigma
_Q}\right] -\arcsin \left[ \frac{S_C}{S_Q}\sqrt{\Sigma _Q}\right]\;,
\label{IntRadn}
\end{equation}
where $Q$ is a reference point in this period, $S_C<S_D\leq S_Q\sqrt{\Sigma
_Q}$. For the early radiation dominated era, we get the corresponding
expression for $\Psi _1$ with $C$,$D$ replaced by $A$,$B$ respectively, and $ 
S_C\leq S_D$. For the inflationary era, we set $\gamma =0$ and obtain

\begin{equation}
\Psi _2\equiv \Psi (0,B,C)=\arccos \left[ \frac{S_C}{S_P}\sqrt{\Sigma _P}
\right] -\arccos \left[ \frac{S_B}{S_P}\sqrt{\Sigma _P}\right]\;,  
\label{IntCC}
\end{equation}
where $P$ is a reference point in this period, and $S_P\sqrt{\Sigma _P}\leq
S_B\leq S_C.$ An alternative expression in the latter case may be obtained
by integrating the second integral in (\ref{Psi}) with scale factor 
(\ref{CC}). The result is

\begin{equation}
\Psi _2\equiv \Psi (0,B,C)=2\left( \arctan \left[ \exp (\lambda t_C)\right]
-\arctan \left[ \exp (\lambda t_B)\right] \right)  \label{IntCC_alt}
\end{equation}
where $t$ is given in terms of $S$ by (\ref{t_i}), and $t_C>t_B\geq 0$.

From these expressions follow the causal limits

\begin{equation}
\;\Psi _1<\pi /2,\;\Psi _2<\pi /2,\;\Psi _3<\pi /2,\;\Psi _4<\pi
\label{Limits}
\end{equation}
for the various epochs when the universe is always expanding. Including the
collapse phases would double the limits. Hence when $k=+1$, no matter how
long inflation lasts\footnote{ 
As mentioned above, in this and the following sections we set $\Psi
_{Planck}=0\ $.}, 
\begin{equation}
\;\Psi _H(t_f)\leq \pi ,\;\Psi _H(t_{eq})\leq 3\pi /2,\;\Psi _H(t_0)\leq
5\pi /2.\;  \label{Limit1}
\end{equation}
One can modify this in obvious ways for alternative inflationary scenarios.

\subsection{Integration Results}

Detailed integration gives much stronger results. Defining constants $c_i$
by 
\begin{equation}
\;\;\Psi _i=c_i\frac \pi 2,\   \label{Def_c}
\end{equation}
we get the following estimates for the late radiation era and matter
dominated era, using current data and the Friedman equation: 
\begin{equation}
c_3\approx 0.0002,\;c_4\approx 0.12.  \label{Est_c}
\end{equation}
The quantity $c_4$ (corresponding to the visual horizon size) is small
because the universe is nowhere near recollapsing at present; while $c_3$
(corresponding to the particle horizon in a simple Hot Big Bang model) is
even smaller:\ $c_3/\;c_4\approx $ $0.0002/0.12=$ $1.67\times 10^{-3},$
which is just the usual result that there is indeed a major horizon problem
in the standard Hot Big Bang model.

To estimate the inflationary era contribution $c_2$, assume N e-foldings,
where $N\geq 60$ : then $S(t_C)\;=e^NS(t_B)$. The extreme case is a universe
started from rest, with $t_B=0$ in expression (\ref{CC}), which implies $ 
\arctan \left[ \exp (\lambda t_B)\right] =\pi /4.$ Then for $N=60$, $\arctan
\left[ \exp (\lambda t_C)\right] =\arctan \left[ 2\exp 60\right] $ $\approx
\pi /2,$ so (\ref{IntCC_alt}) gives $\Psi _2\approx 2\left( \frac \pi
2-\frac \pi 4\right) =\frac \pi 2,$ the maximum allowed value. For any other
allowed case, $t_B>0$ (see (\ref{t_i})) and the initial expansion rate is
non-zero. For the same number of e-foldings, the $t_C$ term will remain at
effectively the same value (the $\arctan $ curve being essentially vertical
for these values), but the $t_B$ term can take any value less than $\frac \pi
2,$ depending on the chosen value of $t_B$, or equivalently, the initial
value of $\Omega _\Lambda (t_B)$. Indeed $\ t_B\ =\frac 1\lambda \arctan h 
\sqrt{\frac 1{\Omega _\Lambda (t_B)}}$ which can take any value $ 
0<t_B<\infty $ as $\Omega _\Lambda (t_B)$ ranges over its allowed values $ 
1<\Omega _\Lambda (t_B)\leq \infty .$ Consequently as $t_B$ varies, $\pi
/4\leq \arctan \left[ \exp (\lambda t_B)\right] <\pi /2,$ and for the same
number of e-foldings, $\Psi _2$ (given by (\ref{IntCC_alt})) can take any
value from approximately zero\ $(c_2=0)$ to $\pi /2\;(c_2=1).$ Given any
specific choice for $t_B$, increasing the number of e-foldings (and so
bringing the final value of $\Omega _0$ closer to unity) will make no
difference to this outcome: the first term in equation (\ref{IntCC_alt}) 
has already
reached its limit for all practical purposes, and any further increase in $ 
t_C$ makes no difference. The key point is how close to the limiting value
of infinity the initial value of $\Omega _\Lambda $ is, that is, how close
to stationary the start is. If it is not close to that limit, then the value
obtained for the integral will be small.

It turns out that as a consequence of the junction conditions between the
first radiation era and the inflationary era, $c_1\approx c_2.$ Thus when $ 
k=+1$, no matter how long inflation lasts, on using (\ref{Est_c}) 
\begin{equation}
\Psi _H(t_{eq})=(c_1+c_2+c_3)\frac \pi 2\approx 2c_2\frac \pi 2=\;\Psi
_H(t_f),\;0<c_2\leq 1.\   \label{Ests}
\end{equation}
It follows that zero-rest mass radiation traveling freely can at most just
manage to circle the universe once before decoupling, no matter how much
inflation there is, because $\Psi$ increases at most by $\pi$ in each
direction before decoupling. Thus the kind of multiple particle exchange
that would be needed to set up similar conditions over the entire LSS is
simply not possible.

\section{The Horizon Problem}

To examine the horizon problem for the CBR, we need to consider causal
relations at decoupling. These depend on two length scales on the LSS (given
by $t=t_{eq}$, $z_{eq}=1200)$: the sizes $D_H(t_{eq})$ of the particle
horizon and $D_{VH}(t_{eq})$ of the visual horizon at that time, determined
respectively by

\begin{equation}
D_H(t_{eq})=S(t_{eq})\Psi _H(t_{eq}),\;D_{VH}(t_{eq})=S(t_{eq})\Psi _4\;,
\label{AngSize}
\end{equation}
with $\Psi _H(t_{eq})$ given by (\ref{Horiz_teq}) together with (\ref
{IntRadn},\ref{IntCC}) and $\Psi _4$ given by (\ref{IntDust}). Three cases
can arise, given that we know from the above estimates that $\pi >2\Psi
_{VH}(t_{eq})$.

\textbf{Case 1}. $\Psi _H(t_{eq})\geq \pi >\;2\Psi _{VH}(t_{eq}):$ all
points on the LSS are in causal contact and their combined images cover the
entire sky; thus the horizon problem is solved on all angular scales (see
Fig. 1). There are no event horizons by the end of inflation.

While this can happen -- when there is an extremely high value for
$\Omega_{\Lambda}$ (much, much larger than $1$) -- (\ref{Ests}) together
with realistic estimates for $c_2$ shows this is not true in 
most inflationary universes when $k=+1$. So we need to consider 
the situation when $\Psi _H(t_{eq})<\pi $. The geometry
of the situation then is as follows: the visual horizon corresponds to the
intersection of our past light-cone $C^{-}(E)$ with the LSS ($E$ is the
point here and now) , which is a 2-sphere $\mathcal{C}_H$ of radius $ 
D_{VH}(t_{eq})$ in the LSS, centered on our past world line $\gamma $. The
particle horizon of any point $q$ in the LSS is a 2-sphere $\mathcal{S}_H$
of radius $D_H(t_{eq})$ in the LSS, centered on $q$, generated by the
creation light cone of the observer \footnote{ 
Often people define the particle horizon as the set of world lines emanating
from the points of the initial singularity where our past light cone
intersects it, see e. g. Kolb and Turner, \textit{The Early Universe} \cite
{Ref1}. Clearly the definition we are using here is equivalent to that.}.
When $q$ is on $\gamma ,$ these two spheres will be concentric. Now two
cases are possible.

\textbf{Case 2.} $\pi >\Psi _H(t_{eq})>2\Psi _{VH}(t_{eq}):$ The horizon
problem is solved in a theoretical sense when at least one photon or
graviton can be interchanged between each observable point on the LSS (see
Fig. 2). This will be the case if $\Psi _H(t_{eq})>2\Psi _{VH}(t_{eq})=2\Psi
_4,$ the factor 2 arising because we demand that points on the LSS that we
see in opposite directions in the sky are able to communicate with each
other (Note that these points are unable to communicate with each other, 
because $c_4<\pi /2.)$. Thus from (\ref{Est_c},\ref{Ests}), the 
requirement for solving the horizon problem in this sense is 
\begin{equation}
\Psi _H(t_{eq})>\;0.24\frac \pi 2\Leftrightarrow (c_1+c_2+c_3)\approx
2c_2>0.24\;.  \label{Solve}
\end{equation}
This is possible for some $k=+1$ inflationary models, as we see by the above
analysis.\emph{\ }

\textbf{Case 3.} $\pi >2\Psi _{VH}(t_{eq})>\Psi _H(t_{eq}):$ The horizon
problem is not solved in the sense that points on the LSS that we can
observe (they lie within the visual horizon) are not pairwise-causally
connected to each other (see Fig. 3). This will be the case when 
(\ref{Solve}) is not true, that is, when 
\begin{equation}
c_2<0.12.  \label{NotSolve}
\end{equation}
There are many $k=+1$ inflationary models for which this is true,
irrespective of how many e-foldings occur; they simply have to start well
away from the minimum of cosh$\lambda t$ (see (\ref{CC})), which is given
by $\Omega_{\Lambda} = \infty$.

\subsection{Causal Diagrams}

How is this related to the usual causal diagrams \cite{Ref4a}, that suggest
that the horizon problem is solved by inflation pushing down the start of
inflation $t=t_i$ arbitrarily far in those diagrams when inflation occurs? 
\cite{Horiz}. The point here is that when $k=+1$, we can't push the initial
surface $t=t_i$ down arbitrarily far in those diagrams, however much
inflation is allowed, because the integral (\ref{Int}) is bounded, see (\ref
{Limits}) above and the conformal diagrams in (\cite{Ref4a}). In analytic
terms, the difference is essentially that between evaluating this integral
for a de Sitter universe in a $k=0$ ($a(t)\;=expHt$) frame, when we can push
the integral back to negative times as far as we like (the epoch $t=0$ being
arbitrarily assigned), and evaluating it in a $k=+1$ ($a(t)=\cosh Ht)$ frame 
\cite{Schr}, where the time $t=0$ is a preferred time (the turn-around time
for the scale factor) and is the maximum to which the integral can be
extended \cite{Paper_1}, see ( \ref{t_i}). The integral $\int dt/a(t)$ is
quite different in these two cases, indeed \emph{this integral has a
discontinuous limit as the spatial curvature } $K(t_o)=k/S^2(t_o)$ \emph{ 
goes to zero}: for $\Omega \leq 1$, it is unbounded with unbounded
integration limits, but for all $\Omega >1$, it has bounded integration
limits and is bounded by the limits given above. This is possible because
when $K>0$ and $\Lambda $ is indeed constant (or almost constant) the $k=+1$
term always dominates this integral eventually at early enough times, no
matter how small $K$ is today (until radiation kicks in and becomes the
dominant term: but that is the end of inflation). Thus given any desired
number of e-foldings, evaluating the integral for $\Omega _0=1$ (as is
usually done) gives quite a different result from evaluating it for $\Omega
_0=1+\varepsilon $ with $\varepsilon >0,$ no matter how small $\varepsilon $
is.

\subsection{Realistic Estimates}

In order to estimate how probable cases 2 and 3 are, we have examined
a grid of models of the kind described above in which the inflationary
epoch has at least 60 e-foldings, and is varied by allowing (i)
different starting times after the end of the Planck epoch (i.e.
different periods of radiation domination before inflation commences),
(ii) different ending times well before the nucleosynthesis epoch but
below the GUT energy, and (iii) different final values of the density
parameter $\Omega _0$. Details are given in \cite{Paper_2}. The
conclusion is that \emph{in most cases inflation will not succeed in
solving the horizon problem} because  (\ref{Solve}) is not true. The
only cases in which inflation will solve the  horizon problem are
those in which it begins very close to the turn-around in the $\cosh$
function, that is at a very nearly static state and with an enormously
high value for $\Omega_{\Lambda}$. The essential point is that given
any chosen starting conditions, the integral (10) rapidly comes very
close to its final value and thereafter no matter how much more
inflation takes place, it adds a negligible amount to this integral;

\section{The Relation to Homogeneity}

The above causal analysis gives upper limits to the scales on which causal
processes can operate. But single contact by massless particles is clearly
insufficient to cause homogenization; much more interaction is needed.
Additionally, the effect of interactions restricts realistic causation much
more. The extremely short mean free path for matter and radiation in the
radiation-dominated era implies that only massless neutrinos and
gravitational radiation travel at the speed of light in this epoch, and they
cannot cause homogeneity; massive particles and electromagnetic radiation
travel much slower. Thus effective causal interactions will come from a much
more restricted domain at early times than indicated by causal horizons
based on the local speed of light. This means the true horizon problem is
even greater than indicated by the above estimates.

To examine this in detail, we need an estimation of the domain that causes
significant effects locally in the neighbourhood of our Galaxy, as a
function of time (or equivalently, of scale factor) - how large was this
domain at nucleosynthesis, at baryosynthesis, at the end of inflation, at
the Planck time? There are three major physical effects to consider:
nucleosynthesis, smoothing and structure formation (growth of density
inhomogeneities). A detailed discussion will be given in \cite{Paper_2}.

\subsection{Chemical Composition:\ Uniform Thermal Histories}

The local composition of matter depends on the relevant thermal histories of
that matter, determined by local conditions near the particle world lines in
the early universe. A uniform chemical composition on large scales thus
depends on uniform thermal histories occurring in widely separated regions
in the early universe \cite{Ref9}. The point is that while some diffusion of
elements will take place after nucleosynthesis, this will be strongly damped
by the expansion of the universe; neither particles nor radiation can move
freely because of tight coupling between them, so element abundances set up
early on will tend to stay put in the same (comoving) place. There might
for example be an initial spatial variation in the baryon number density,
lepton number density, and charge density, as well as in local densities and
expansion rates, generating different photon-to-baryon ratios in different
regions, and hence resulting in spatially varying nucleosynthesis patterns.
The resulting inhomogeneous element abundances will remain unchanged in the
same comoving locations until decoupling has taken place and star formation
has begun.

The issue, then, is the particle horizon size at the times of
nucleosynthesis and baryosynthesis, determining the limits of causality for
the epochs of baryosynthesis and nucleosynthesis, and hence for resulting
uniform thermal histories at later times. These scales will be more or less
the same as those at the end of inflation, for which speed-of-light limits
are given by $\Psi _H(t_f),$ estimated above (\ref{Ests}); they will be seen
on the surface of last scattering as in (\ref{Horiz_Dist}). From the
estimates above, the implication is that in most $k=+1$ models, there will
not be a possibility of setting up causally equalized conditions for
nucleosynthesis by physical processes occurring during inflation. The
element abundance sky will consist of many causally disconnected domains.

\subsection{Smoothing by Expansion}

What will take place unchanged is the smoothing out that is associated
directly with expansion, which smooths out structures locally, independent
of what happens at distant places. The argument is simple: choose a smooth
enough domain, however small, at the end of the Planck era; enough
e-foldings during inflation will make it larger than the visual horizon size
at decoupling, and so will explain the observed homogeneity today. This is
what is understood by many as the major causal mechanism by which inflation
causes homogeneity at late times. On this scenario, the large-scale
homogeneity we measure today is due to homogeneity on very small physical
scales being set up prior to inflation, during the Planck epoch, so we can
no longer ignore $\Psi _{Planck},$ as we have done up to now. Whether or not
inflation is able to solve the horizon problem of causal connectivity is
then irrelevant; the necessary homogeneity (on a very small physical scale)
was created before inflation began, and then preserved when one follows the 
comoving evolution of inhomogeneities.

The causal implications for the Planck era are quite severe. Indeed it is
clear from the relevant causal diagrams (see Fig. 4) that the essential
requirement for this to succeed is that 
\begin{equation}
\Psi _{Planck}>\Psi _{VH}=\Psi _4:  \label{Planck}
\end{equation}
at the beginning of inflation, there was already causal connectivity on a
scale larger than the scale of the entire visible universe today retrodicted
to that time. When this is true,
whether or not this local process in fact leads to smoothing is then totally
dependent on (i) what relics are left over from the quantum gravity era at
what wavelengths, and (ii) on how uniform the subsequent expansion is.

\section{The quantum gravity problem}

Ultimately causal estimates depend on unknown physics in the Planck era,
where space-time foam, a lattice domain, or tumbling light cones may
occur and determine causal connectivity at the start of inflation (Guth
emphasizes that the initial size of an inflationary patch need be only one
billionth the size of a single proton \cite{Ref1}.). It is certainly clear
that physics in the Planck era influences initial conditions for inflation,
and hence the anisotropy and inhomogeneity spectra observed today \cite
{transPlanck}; what is not clear is that the almost-FL studies carried out
so far give anything like the correct answer. If conditions are very
inhomogeneous, almost everywhere inflation may not succeed in starting;
however when it does succeed, it will soon dominate the local universe
region in volume terms. In that region, inflation will dramatically amplify
the comoving scales associated with whatever inhomogeneity there is to begin
with\emph{. }The remnants of quantum gravity may not be smooth: they may be
arbitrarily inhomogeneous, even fractal - and usual inflationary studies do
not consider this full range of possibilities \cite{Penrose}. In contrast,
some studies propose quantum mechanism that will indeed create the universe
in a smooth state that solves the homogeneity problem in $k=+1$ models
before inflation ever begins \cite{Linde}, but as the link between quantum
gravity and quantum cosmology models is not yet firmly established, this
proposal must be treated with some caution.

Two principal questions we must therefore address to quantum cosmology are: 
\textbf{1.} What processes in the Planck era were responsible for the
initial causal self-connectedness and homogeneity of the primordial universe
at the Planck transition (whatever its size)? and \textbf{2.} What
determines the limiting size of such a region -- what are the limits to the
correlations quantum gravitational process can establish at the Planck
transition? Even if it turns out that this limit is indeed the Planck
length, the first question demands an adequate answer. And, given that space
and time -- and therefore causality itself -- would not have anywhere near
the same structure in the Planck era as after the transition to classical
space-time, the second question also demands careful consideration.

Thus, the fundamental problem is that we don't know the causal connection
size during the quantum gravity era nor at the Planck time. We can calculate
it in a \ FRW context, but that context will not obtain at very early times
when quantum fluctuations in space-time structure are severe. Nevertheless,
we need to estimate the Planck contribution $\Psi _{Planck}$ in order to
truly understand the range of causality in the later universe. And the
`smoothing by expansion' proposal can only work if (\ref{Planck})\ is
satisfied as a result of those processes.

\section{Event Horizons}

Event horizons \cite{Ref4} occur if $\Psi (E,F)\equiv
\int_{t_E}^{t_F}(c/S(t))dt$ is bounded as $t(F)\rightarrow t_{\max }$. From
the integral (\ref{IntDust}), this is indeed the case when if $k=+1$ and $ 
\Lambda =0$, for then always $\Psi (1,E,F)<\pi $. If there is a cosmological
constant in such models that will only make the situation worse, because
such a constant by itself will always (i.e. with the single exception of the
highly unstable case of a model asymptotic to an Einstein Static universe in
the future) lead to this integral being bounded even if 
$t\rightarrow \infty$. That is, irrespective of the value of a 
cosmological constant, and, whether
or not there is some entity like quintessence present, there will always by
event horizons in $k=+1$ FL\ universe models. Thus the alleged problems for
string theory resulting from the existence of event horizons \cite{Ref5}
will \emph{always} be implied by such models. This is in addition to any
problems arising because the space sections are compact, so that there is no
infinity to use for setting data.

However it is not clear that this would necessarily be a death-knell for
string theory, even if we were eventually to conclude conclusively that $k=+1
$ in the real universe. The key point here is that string theory is in
essence a theory of small scale structure and quantum gravity properties,
and we are here considering properties of the universe on the largest
observable scales, and indeed on scales that might never be observable (c.f. 
\cite{Ref11}). One might suggest that an `effective infinity' for $S$-matrix
calculations for string theory could be at a finite distance from a local
object (for example, at CERN the `effective infinity' where the outermost
measurements are made is at a distance of about 10 meters from where the
particle collisions take place), rather than having to be taken literally to
infinity (which is way outside the visual horizon -- so we have no chance of
knowing what conditions are like there anyhow). Thus it may be worthwhile
pursuing a somewhat more local version of the setting of data for string
theory, in line with the spirit of the `finite infinity' proposal in \cite
{Fi}.

\section{Conclusion}

If the universe has positive spatial curvature $(k=+1)$, then no matter how
much inflation takes place, effective causality since the Planck time is
almost always smaller than the whole LSS -- unless there were extreme
conditions right at the beginning of inflation, that is, no significant cosmic
expansion before that. The CBR intensity sky, mirroring
the density fluctuations at last scattering that later led to structure
formation, will usually consist of causally disconnected regions, and in
these cases the same applies to the element abundance sky, mirroring the
early epoch of nucleosynthesis. If inflation is going to solve the horizon
problem in all cases, we must have $k=-1$ (given that $k=0$ is infinitely
improbable). If $k=+1$ is observationally indicated, that will suggest that
we need (in practically all cases) physical homogeneity prior to inflation, 
because it can be created by physical processes during inflation only in the
extreme case of an enormously large (virtually infinite) $\Omega_{\Lambda}$,
corresponding to virtually no cosmic expansion before inflation itself.
This conclusion in fact accords
with the understanding many have of inflation as simply expanding already
homogenized patches of the universe, smoothed out by processes at work in
the Planck era. While this may be a plausible mechanism, it is somewhat
surprising to see this proposed causal structure (shown in Figure 4), based
on comoving timelike world lines, given the emphasis placed in much of the
inflationary literature on the way that inflation is directly able to solve the
horizon problem.

It should be noted that this conclusion is based purely on examining
inflation in FL\ universe models with a constant vacuum energy, and is not
based on examinations of Trans-Planckian physics on the one hand, nor on
studies of embedding such a FL\ region in a larger region on the other, nor
does it take scalar field dynamics into account. In may be that physics in
the Planck era will smooth things out on large enough comoving scales that
the universe today is spatially homogeneous simply by comoving expansion; in
that case the horizon problem is irrelevant. This indeed appears to be the
option proposed in many inflationary studies. We suggest that in that case,
this position should be made clear. Then inflation is not required to solve
the causal issues raised by the horizon problem; it is the Planck era that
is assumed to do so, despite the unknown physics of that era.

We do not expect any major difference from our results to occur for the case
of more realistic exponential inflationary models of the early universe. But
power-law inflation may give quite different answers, and one could solve
the problem by models that are not slow-rolling for a major part of the
scalar-field dominated early era; but then they fall outside the standard
inflationary paradigm. We are fully aware that in order to properly study
the issue, we need to examine anisotropic and inhomogeneous geometries
rather than just FL models, because analyses based on FL\ models with their
Robertson-Walker geometry cannot be used to analyse very anisotropic or
inhomogeneous cases. Nevertheless this study shows there are major causal
differences in inflationary FL universes with $k=+1$ or $k=-1,$ influencing
the ability of physical processes to causally homogenise the universe. The
implication is (a) that we need to try all observational methods available
to determine which is the case, because this makes a significant difference
not only to the spatial topology, but also to the causal structure of the
universe, and (b) we should examine inhomogeneous inflationary cosmological
models to see if any similar difference exists in those cases between causal
behaviour of models that are necessarily spatially compact, and the rest.
\section*{Acknowledgments}
We thank Roy Maartens, Bruce Bassett, and Claess Uggla for useful comments, 
and the NRF (South Africa) for financial support.

\newpage
\begin{figure}
\includegraphics{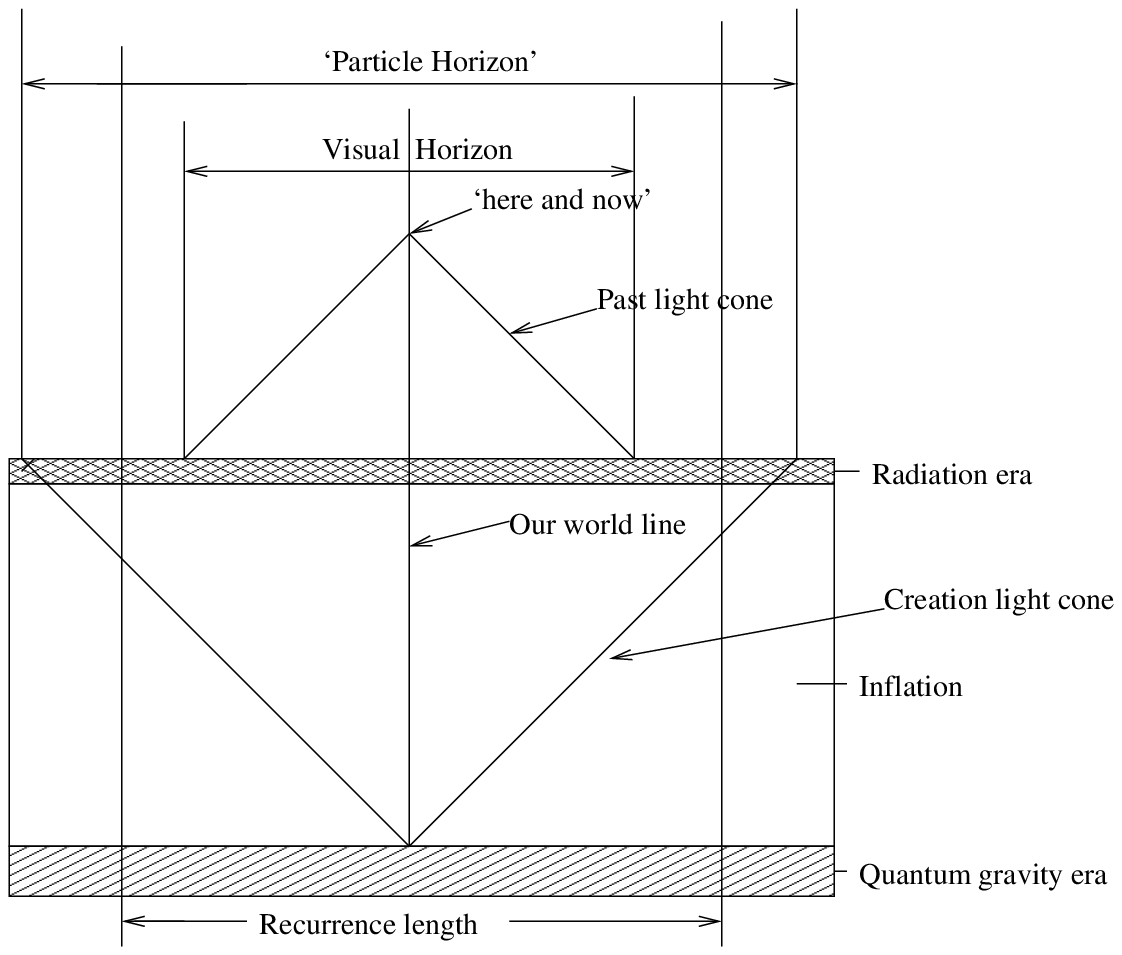}
\caption{Creation lightcone larger than the identification scale. There are
no particle horizons and all events at last scattering are causally 
connected.}
\end{figure}

\begin{figure}
\includegraphics{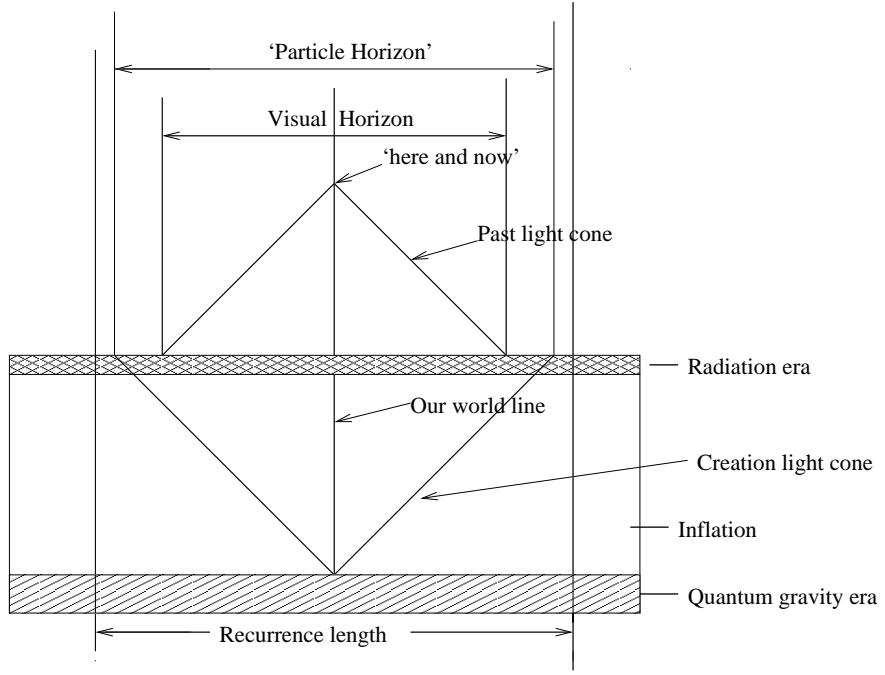}
\caption{Creation lightcone less than the identification scale but larger than the visual horizon. There are particle horizons, but all seen events at last scattering are causally connected.}
\end{figure}

\begin{figure}
\includegraphics{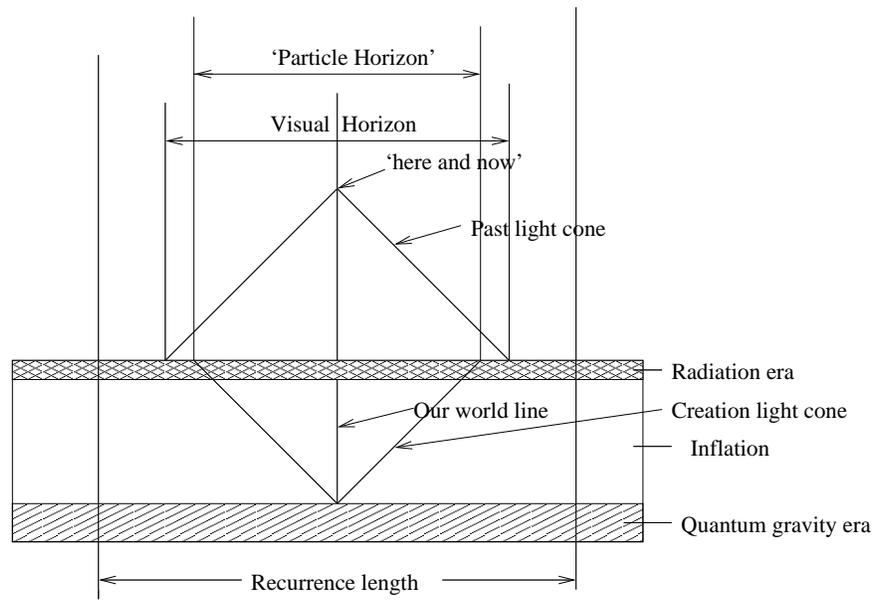}
\caption{Creation light cone less than the identification scale and visual horizon. There are particle horizons, and not all seen events on last scattering are causally connected.}
\end{figure}

\begin{figure}
\includegraphics{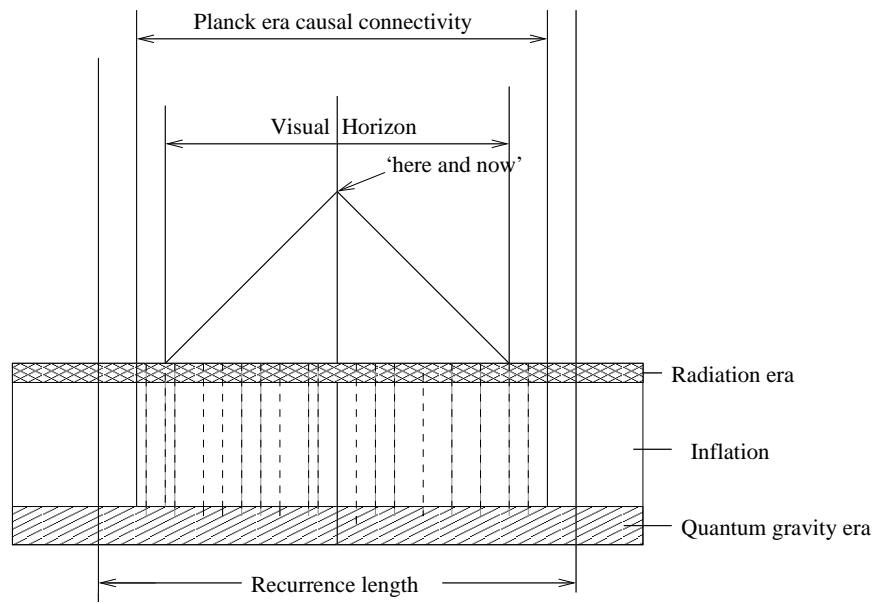}
\caption{Causality generated by comoving expansion of world-lines. The vertical lines are comoving, and causality generated at the Plank scale is assumed to be bigger than the visual horizon. The creation light cone is irrelevant.}
\end{figure}

\end{document}